\documentclass[%
 reprint,
 amsmath,amssymb,
 aps,
]{revtex4-2}
\usepackage{chemformula}
\usepackage{graphicx}% Include figure files
\usepackage{dcolumn}% Align table columns on decimal point
\usepackage{bm}% bold math
\usepackage{amsmath}
\usepackage[version=4]{mhchem}

\begin{document}

\preprint{APS/123-QED}

\title{Non-destructive inelastic recoil spectroscopy of a single molecular ion: a versatile tool toward precision action spectroscopy}% Force line breaks with \\%

\author{Aaron Calvin}
\affiliation{Department of Physics, University of California, Santa Barbara}%
\author{Scott Eierman}
\affiliation{Department of Physics, University of California, Santa Barbara}
\author{Zeyun Peng}
\affiliation{Department of Physics, University of California, Santa Barbara}
\author{Merrell Brzeczek}
\affiliation{Department of Physics, University of California, Santa Barbara}
\author{Samuel Kresch}
\affiliation{Department of Physics, University of California, Santa Barbara}

\author{Elijah Lane}
\affiliation{Department of Physics, University of California, Santa Barbara}

\author{Lincoln Satterthwaite}
\affiliation{Department of Physics, University of California, Santa Barbara}

\author{David Patterson}
\affiliation{Department of Physics, University of California, Santa Barbara}
\email{davepatterson@ucsb.edu}

\date{\today}% It is always \today, today,

% ABSTRACT MUST BE CAPPED AT 600 CHARACTERS
\begin{abstract}
We demonstrate a novel single molecule technique that is compatible with high precision measurements and obtain the spectrum of two molecular ion species. While the current result yields modest spectral resolution due to a broad light source, we expect the method to ultimately provide resolution comparable to quantum logic methods with significantly less stringent requirements.  Adaptations of this technique will prove useful in a wide range of precision spectroscopy arenas including the search for parity violating effects in chiral molecules.

\end{abstract}

\maketitle

%\tableofcontents

Ion traps present a powerful platform for precision spectroscopy of molecular ions. In particular, molecular spectroscopy in mixed-species Coulomb crystals, composed of a molecular ion and laser-cooled atomic ion, affords a level of exceptional molecular control that in turn leads to long interrogation times\cite{Khanyile2015} and unparalleled spectroscopic resolution\cite{chou2020frequency}.  These techniques are actively investigated for applications in testing fundamental physics\cite{Alighanbari2020}, \textit{ab initio} theory\cite{Germann2014}, cold reactions\cite{Heazlewood2019, Krohn2022}, and chemical dynamics\cite{Drfler2019, Puri2019, Venkataramanababu2022}. Within this framework, non-destructive quantum logic spectroscopy (QLS) techniques have emerged as the gold standard for precision spectroscopy of a single molecular ion \cite{Sinhal2023}. These techniques coherently drive the molecule-atom pair translational motion in the trap conditional on the state of the molecular ion and provides a nondestructive amplification of a transition measurement. Despite significant strides in recent years, quantum logic techniques remain technically challenging and have yet to be implemented on a polyatomic molecule.

In contrast, action spectroscopy is a versatile and diverse framework for obtaining the spectra of molecular ions ranging from simple diatomics to large biomolecules\cite{Pereverzev2022}. Tagging spectroscopy, a form of action spectroscopy, weakly adheres an inert gas ``tag'' such as a noble gas atom or N$_2$ via van der Waals interaction with the molecule. Resonant excitation of vibrational modes of the molecule detaches the tag and is detected as a change in mass of the complex\cite{johnson2013review}. Action spectroscopy in laser-cooled Coulomb crystals has been largely limited to destructive photodissociation measurements \cite{Calvin2018, Lien2014}, although a tagging spectrum of single molecular ions was recently demonstrated\cite{Calvin2023}. The lifetime limited resolution of tagging and photodissociation methods have led searches for less perturbative methods\cite{Asvany2021, Asvany2015}.  Schmid\textit{ et al.} recently demonstrated Leak Out Spectroscopy, a method which has been used to record high-resolution ro-vibrational spectra of a wide range of molecular ions\cite{Schmid2022, Asvany2023}.  The technique takes advantage of inelastic collisions between vibrationally excited molecular ions and cold buffer gas atoms or molecules.  These collisions internally cool the molecular ions while simultaneously imparting a large ($\sim$10 meV) kinetic energy recoil to the molecule.  This kinetic energy is sufficient to eject the molecular ion from the trap, where it is detected.  This method yields a spectral resolution of $\sim$MHz, set by the first-order Doppler width of the cold ($\sim$15 K) molecular ions.  Our method extends Leak Out Spectroscopy to non-destructively record spectra of single, Coulomb-crystallized molecular ions.

We demonstrate this inelastic recoil spectroscopy (IRS) and record the vibrational spectra of a single tropylium (Tr$^+$, C$_{7}$H$_{7}^{+}$) molecule and a single benzodioxol fragment (Bd$^+$, C$_{7}$H$_{5}$O$_{2}^{+}$) molecule. The molecule of interest is co-trapped with laser-cooled $^{88}$Sr$^{+}$ atoms, and inelastic collisions are detected non-destructively via observation of the $^{88}$Sr$^{+}$ fluorescence. Thus, our method retains the generality of the Leak Out Spectroscopy technique, but allows for spectroscopy to be performed on highly-controlled single molecular ions, as in QLS.  Future experiments can be performed in the Lamb-Dicke regime, which is immune to first order Doppler shifts.  This combination of generality and precision suggests the method will represent a significant contribution to the molecular spectroscopy tool suite.

%%%%%%%%%%%%%%%%%%%%%%%%%%%%%%%%%%%%%%%%%%%%%%%%%%%%%%%%%%%%%%%%%%%%%%%%%%%%%%%%%%%%%
%\section{\label{sec:level1}EXPERIMENTAL}
%\subsection{\label{sec:level2}Construction of the double well trap}
%%%%%%%%%%%%%%%%%%%%%%%%%%%%%%%%%%%%%%%%%%%%%%%%%%%%%%%%%%%%%%%%%%%%%%%%%%%%%%%%%%%%%

Our apparatus is a slightly modified version of the apparatus used to obtain the tagging spectrum of single molecular ions \cite{Calvin2023, Eirmann2023}, shown in Fig. 1 (a).  A linear quadrupole Paul trap is enclosed in a 7 K copper shield, and can be mass-selectively loaded with $^{88}$Sr$^{+}$ ions and molecular ions as in \cite{Eirmann2023}.  We load a single molecular ion and a small number (1-3) of $^{88}$Sr$^{+}$ atoms. The mixed-species ensemble is laser cooled into a Coulomb crystal via Doppler-cooling on the $5S_{1/2} \rightarrow 5P_{1/2}$ 422 nm transition in $^{88}$Sr$^{+}$ (Doppler temperature $\sim$ 300 $\mu$K).  The ensemble is exposed to low pressure He gas at $\sim$8 K. Based on the time ($\sim$100 ms) between $^{88}$Sr$^{+}$ - molc.$^+$ $ \rightarrow $ molc.$^+$ - $^{88}$Sr$^{+}$ reorganization events, we estimate a He pressure on the order of 10$^{-8}$ torr \cite{Eirmann2023}. This estimate is consistent with the a base pressure reading of 3 $\times$ 10$^{-8}$ torr from an ion gauge that measures the pressure of the room temperature dewar that contains the cryo-cooled trap. Collisions with this gas serve to rotationally cool the molecule, but are not sufficient to melt the Coulomb crystal.  C-H stretch modes of the molecular ion are excited with a mid-IR (3030 cm$^{-1}$) OPO.  As in references \cite{Schmid2022, Asvany2023}, we rely on inelastic collisions of a vibrationally excited molecule with a cold (8K) buffer gas particle, typically He. For a vibrational excitation of Tr$^+$ at 3030 cm$^{-1}$, a quenching resulting from the inelastic collision distributes the 375 meV (4350 K) between the vibrational and rotational degrees of freedom for the molecule as well as translational energy for both collision partners. A kinetic energy of about 16 meV (183K) imparted on the molecular ion is enough to substantially reorganize the configuration of atomic and molecular ions in the trap whereas elastic collisions of 8 K He do not. The relatively large energy scales of the inelastic recoil makes ground state cooling unnecessary, at the cost of quantum state preparation as is realized in QLS. The locations of the molecular ion are inferred from the position of the atomic ions imaged using laser-induced fluorescence. 

We demonstrate IRS in two distinct configurations.  In the \emph{double well configuration}, the molecular ion is held in one of two wells, with a barrier chosen such that the inelastic recoil is sufficient to cross between two trapping regions.  In the \emph{single well configuration}, the inelastic collision momentarily knocks the molecular ion out of a two-ion crystallized configuration, leaving the atom trapped in the trap center. The ion positions are monitored by an EMCCD (Q Imaging Retiga R1) from both the side and from below the ion trap. A portion (70\%) of the fluorescence from the bottom is sent to a PMT to enable the measurement of the axial secular frequency via Doppler-modulated fluorescence.\cite{Eirmann2023} The axial secular frequency is used to both confirm the mass of a co-trapped molecular ion \cite{Kielpinski2000} and determine the distance between ions \cite{James1998}. 

%%%%%%%%%%%%%%%%%%%%%%%%%%%%%%%%%%%%%%%%%%%%%%%%%%%%%%%%%%%%%%%%%%%%%%%%%%%%%%%%%%%%%
%\subsection{double well configuration}
%%%%%%%%%%%%%%%%%%%%%%%%%%%%%%%%%%%%%%%%%%%%%%%%%%%%%%%%%%%%%%%%%%%%%%%%%%%%%%%%%%%%%

To realize the double-well trap shown in Fig. 1, an additional center pin electrode was added 4 mm above the trap center. When no voltage is applied to this center pin, the trap has a single well. Both the desired molecular ion and $^{88}$Sr$^{+}$ are trapped initially in the single well configuration. A voltage is applied to the center pin electrode and the trap is re-compensated \cite{Saito2021} to produce the double-well potential. We typically observe bifurcation around 65 V on the center pin electrode. The center pin voltage is raised further to 75-80 V to minimize the background flipping events where ions change position between the left and right well. We find this barrier height still suitably low enough to allow the molecule to flip between wells during inelastic collisions. We also find that the trap should be reasonably well compensated to prevent background flipping events, presumably during energetic collisions activated by the micromotion.  The axial double well potential, determined by the locations of the minima and the secular frequencies within each well, is shown in Fig. 1.

\begin{figure}
    \centering
    \includegraphics[width = 0.5\textwidth]{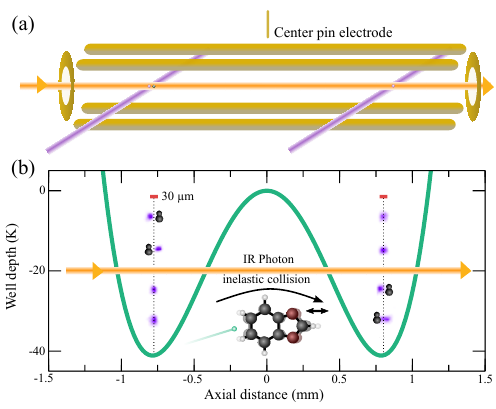}
    \caption{(a) A schematic of the trap (not to scale) shows the layout of the center pin relative to the rest of the linear Paul trap. Details of the trap are described in \cite{Eirmann2023}. Two radial cooling beams (violet) are aligned to laser cool $^{88}$Sr$^{+}$ in each well. The mid-IR OPO (orange) is aligned along the trap axis with an axial cooling beam (not shown). (b) The estimated double well potential along the axial direction of the trap is shown. The false-colored violet spots represent $^{88}$Sr$^+$ ions isolated from a series of four frames taken with a camera imaging both wells. The molecule is represented by a space-filling cartoon with a dotted guide added to highlight the single ion trap position for each well. In this case, a Bd$^+$ molecular ion flips to the right well between the second and third frames, following a 500 ms exposure to the mid-IR OPO on resonance with a dipole-allowed C-H stretch mode and an inelastic collision with He. The 30 $\mu$m distance between ions in the frame is estimated from the measured axial secular frequency.}
    \label{fig:flippytoon}
\end{figure}

We estimate the barrier height around 40 $\pm$ 10 K.  This height is large compared to the laser-cooled motional temperature of the mixed-species crystal, but small compared to the recoil energy from a vibrational-quenching transition.  Without laser cooling, the ion temperature quickly warms to $\sim$8K via elastic collisions with background gas, which is sufficient to reshuffle the ions between wells reasonably quickly.  We indeed observe that both atomic and molecular ions reshuffle between the two wells at random when the cooling light is momentarily blocked.

During an experimental cycle, a single molecular ion is co-trapped with one or two atomic ions in the left well of the double trap. The atomic ions provide sympathetic cooling to millikelvin temperatures as well as providing a visual confirmation of the presence of the molecular ion. A single $^{88}$Sr$^{+}$ is trapped in the right trapping region to provide confirmation of a flip event. A mechanical shutter opens to expose the molecular ion to a mid-IR OPO beam co-aligned with the axial cooling beam. The total exposure time before the molecular ion flips to the right potential well is recorded and used to estimate a flipping rate. A total of 6 data points at each wavelength step was used to obtain the spectra in Fig. 2. After recording a molecule flipping event, the configuration is reshuffled as necessary.

\begin{figure}
    \centering
    \includegraphics{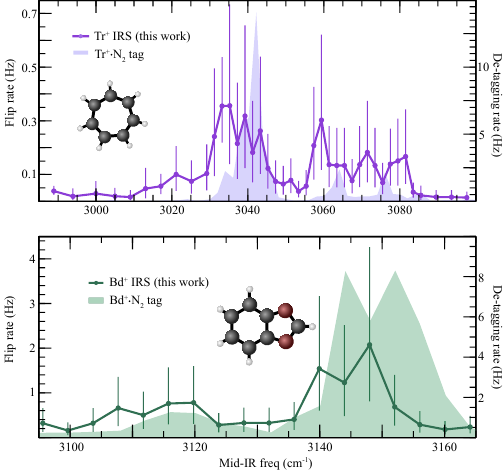}
    \caption{The inelastic recoil spectrum for Tr$^{+}$ and Bd$^{+}$ are shown with a maximum likelihood estimate of the rate with error bars at the 95\% confidence interval (see \cite{Calvin2023, Eirmann2023}).  Both spectra are in agreement with the spectra taken by tagging the respective molecule shifted by 4.5 cm$^{-1}$. Error bars on the tagging spectra are suppressed for clarity. }
    \label{fig:flipflopmodes}
\end{figure}

%%%%%%%%%%%%%%%%%%%%%%%%%%%%%%%%%%%%%%%%%%%%%%%%%%%%%%%%%%%%%%%%%%%%%%%%%%%%%%%%%%%%%
%\subsection{single well configuration}
%%%%%%%%%%%%%%%%%%%%%%%%%%%%%%%%%%%%%%%%%%%%%%%%%%%%%%%%%%%%%%%%%%%%%%%%%%%%%%%%%%%%%

The spectra in Fig. 2 demonstrate the double well configuration of IRS with two molecular ions previously studied using tagging spectroscopy with N$_2$ \cite{Calvin2023, Eirmann2023}. The tag perturbs vibrational energies on the order of a few cm$^{-1}$ \cite{Tsybizova2020, Johnson2014}, with the shift observed in this IRS spectrum compared to previous tagging experiments indicating this method does measure the bare molecular ion. The lifetime limit for the van der Waals complex used in tagging spectroscopy imposes resolution limit from $\sim$10 GHz to 10 THz \cite{Puttkamer1983}. Without this large lifetime limit, IRS should fully rotationally resolve molecular spectra with a suitably narrow mid-IR source. 

Although the double well method robustly produces spectra, a single well measurement is preferable as a more straightforward approach with a faster data acquisition cycle. As a proof of concept, we implement one possible strategy using a camera to capture inelastic recoil events. As shown in Figure 3, an event occurs when an inelastic collision knocks the molecular ion out of the trap null. For a brief time, the  $^{88}$Sr$^{+}$ returns to the single ion equilibrium position until the molecular ion sympathetically re-cools into the crystal. An increased ion separation increases the re-cooling time (typically observed at a few hundred ms), so the endcap voltages are kept relatively low. In practice, if the endcaps are lowered sufficiently to realize a potential with large ion separation, inelastic collisions knock the molecular ion completely out of the trap.
In order to prevent sample loss while maintaining a large ion separation, the endcaps are kept at a modest potential and the center pin electrode is used to add an anharmonic contribution to the axial potential. A mass-dependent parametric tickle of the molecule's radial secular frequency could also be used to amplify the excursion of the molecular ion from the trap null caused by a recoil event\cite{Schmidt2020} and slow re-cooling for easier detection. The parametric trickle was not applied for the scan in Figure 3 due to the similarity in mass of Tr$^{+}$ to $^{88}$Sr$^{+}$.

Each data point in Fig. 3 requires only 20 seconds of data acquisition per point and already shows consistency with the measured Tr$^{+}$ resonance. Increased signal-to noise is easily achieved with a longer acquisition time in order to build statistics with discrete events. Refinements of this method could meet the demanding data throughput necessary for a fully resolved rotational spectrum of a general polyatomic molecule, with a data rate of $\gtrsim$ 10 recorded event per second.

%%%%%%%%%%%%%%%%%%%%%%%%%%%%%%%%%%%%%%%%%%%%%%%%%%%%%%%%%%%%%%%%%%%%%%%%%%%%%%%%%%%%%
%\section{Applications}
%\subsection{Precision measurement}
%%%%%%%%%%%%%%%%%%%%%%%%%%%%%%%%%%%%%%%%%%%%%%%%%%%%%%%%%%%%%%%%%%%%%%%%%%%%%%%%%%%%%

The low motional temperature and highly controlled environment of a single trapped ion presents an attractive staging ground for precision measurement. In particular, tightly confined ions can be interrogated in the Lamb-Dicke regime\cite{Georgescu2021}, which is free of the first-order Doppler effect\cite{ido2003recoil,alighanbari2018rotational}.  The work presented here was done with a rather low resolution ($\sim$ 6 cm$^{-1}$, 0.18 THz linewidth) widely tunable light source, but the resolution of a future experiment with a narrow spectroscopy laser would be limited only by the natural linewidth of the probed transition, the second-order Doppler effect, and collisional broadening. All contributions are estimated at $\lesssim$ 100 Hz.  Such an experiment would easily resolve rotational lines.  In a rotationally resolved experiment, background gas collisions would cycle the molecule through rotational states until it lands in the probed state and is subsequently vibrationally excited\cite{Schmid2022}.  

The ability to resolve rotational lines, in combination with microwave-mediated enantiomer-specific state transfer, would enable the non-destructive determination of the chirality of a single, trapped molecular ion\cite{eibenberger2017enantiomer, patterson20133wm, satterthwaite20193wm}. Chiral molecular ions are proposed as promising candidates for observation of small parity non-conservation (PNC) effects\cite{Eduardus2023}. An extension of the method in this Letter toward ultra-high resolution spectroscopy with a suitably narrow laser source \cite{argence2015QCL} presents a promising avenue for observing the effect in molecules for the first time, with no requirement to synthesize an enantiopure sample.  

In addition to the possibility of measuring PNC effects, the method presents applications in studying molecules of astrophysical interest. Molecular ions are suspected to play an important role in astrochemistry, but the low density of molecular ions makes obtaining a spectra with sufficient resolution challenging. The enhanced sensitivity of action techniques like tagging have resulted in the heralded identification of C$_{60}^+$ in the interstellar medium \cite{Campbell2015}. Tagging spectroscopy, even with an ``innocent'' tag like He, can shift vibrational spectra of the observed complex relative to the bare molecule and requires careful understanding of these perturbations \cite{Roithov2016}. A spectroscopic method that measures the bare molecular ion is therefor invaluable for astrophysical determination. The long lifetime of a molecular ion in the trap may allow the creation of understudied reactive species via chemical or photofragmentation processes with low formation rates to compliment progress toward this goal\cite{Schmid2022}. Further purification to remove different chemical or isomeric species' contributions to the spectra is not required due to the innate purity of a single molecule.   

Our method can detect vibrational excitation events on a time scale of $\lesssim$ 0.1 second.  This suggests that a spectrum of sufficient resolution to definitively identify single molecules could be recorded in $\lesssim$ 100 seconds.  This makes it an intriguing method for analysis of extremely costly samples, such as samples of molecules of possible biological origin on Enceladus\cite{cassiniMassSpec}, Titan \cite{Cui2009, Ali2015}, and Europa. 

\begin{figure}
    \centering
    \includegraphics{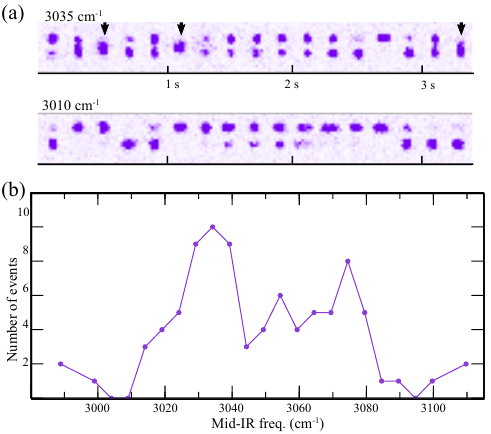}
    \caption{(a) Collated frames (200 ms each) of Tr$^{+}$ and $^{88}$Sr$^{+}$ exposed to mid-IR on resonance (3035 cm$^{-1}$) and off resonance (3010 cm$^{-1}$) of the main C-H stretch transition. On resonance, a number of recoil events knocks Tr$^{+}$ out of the Coulomb crystal and leaves the $^{88}$Sr$^{+}$ in the single ion position as indicated by the arrows. No events were observed in this sequence off resonance. The inter-ion spacing is $\sim$70$\mu$m. (b) The spectrum was recorded as the number of events per 20 second recording at each wavelength step. }
    \label{fig:flipflopmodes2}
\end{figure}

%\section{Discussion and Outlook}

  In conclusion, we have demonstrated a novel, general method for performing vibrational spectroscopy on single molecular ions.  Straightforward extensions of the technique are expected to yield ultra-precise molecular spectroscopy, single molecule chiral readout, and an unambiguous single molecule general chemical analysis method.

 \vspace{5mm} 
%%%%%%%%%%%%%%%%%%%%%%%%%%%%%%%%%%%%%%%%%%%%%%%%%%%%%%%%%%%%%%%%%%%%%%%%%%%%%%%%%%%%%
%\section{Acknowledgments}
%%%%%%%%%%%%%%%%%%%%%%%%%%%%%%%%%%%%%%%%%%%%%%%%%%%%%%%%%%%%%%%%%%%%%%%%%%%%%%%%%%%%%
This work has been supported by the US National Science Foundation (NSF CHE-1912105) and the Air Force Office of Scientific Research (MURI FA9550-20-1-0323).

\end{document}